\documentclass[aps,prl,twocolumn,superscriptaddress,showpacs]{revtex4}
\usepackage{graphicx}
\usepackage{amssymb}
\usepackage{amsfonts}
\usepackage{bm}
\begin{document}

\title{Hole states in wide band-gap diluted magnetic semiconductors and oxides}

\author{Tomasz Dietl}
\affiliation{Institute of Physics, Polish Academy of Science,
al.~Lotnik\'ow 32/46, PL 02-668 Warszawa, Poland}
\affiliation{Institute of Theoretical Physics, Warsaw University,
PL 00-681 Warszawa, Poland} \affiliation{ERATO Semiconductor
Spintronics Project, Japan Science and Technology Agency,
al.~Lotnik\'ow 32/46, PL 02-668 Warszawa, Poland}

\begin{abstract}
Puzzling disagreement between photoemission and optical findings
in magnetically doped GaN and ZnO is explained  within a generalized alloy theory.
The obtained results show that the strong coupling between valence-band holes
and localized spins gives rise to a
mid-gap Zhang-Rice-like state, to a sign reversal of the apparent p-d exchange integral,
and to an increase of the band-gap with the magnetic ion concentration.
\end{abstract}


\maketitle

Recent progress in the understanding of carrier-controlled ferromagnetic
semiconductors such as Ga$_{1-x}$Mn$_x$As  and
p-type Cd$_{1-x}$Mn$_x$Te \cite{Dietl:2006_a} relies to a large extend on comprehensive
optical, magnetooptical,  photoemission, and x-ray absorption (XAS) studies,
which provide mutually consistent information on exchange couplings
between the band states and the localized spins as well as on the positions of d-bands
with respect to the band edges. However, it becomes more and more apparent that this clear cut picture breaks down
entirely in the case of nitride diluted magnetic semiconductors (DMS) and
diluted magnetic oxides (DMO). Here, according to photoemission \cite{Mizokawa:2002_a,Hwang:2005_a}
and XAS \cite{Okabayashi:2004_a} works, the relevant Mn$^{2+}$/Mn$^{3+}$ level resides
in the valence band in both GaN \cite{Hwang:2005_a} and ZnO \cite{Mizokawa:2002_a,Okabayashi:2004_a},
similarly to the case of arsenides, tellurides, and selenides.
Furthermore, the evaluated magnitude of the p-d exchange energy $\beta N_0$ shows the expected enhancement
(due to a stronger p-d hybridization \cite{Blinowski:2002_a} in nitrides
and oxides) to the values of  $-1.6$~eV in Ga$_{1-x}$Mn$_x$N \cite{Hwang:2005_a}
and to $-2.7$~eV \cite{Mizokawa:2002_a} or $-3$~eV \cite{Okabayashi:2004_a} in Zn$_{1-x}$Mn$_x$O,
significantly over the value of $-1$~eV, typical for other DMS \cite{Mizokawa:2002_a,Hwang:2005_a,Okabayashi:2004_a}.
By contrast, optical and magnetooptical studies appear to reveal the presence
of Mn$^{2+}$/Mn$^{3+}$  level in the band gap of both GaN \cite{Wolos:2004_a}
and ZnO \cite{Kittilstved:2006_a}. Moreover, the determined values of $\beta N_0$ show
either opposite sign and/or much reduced amplitude comparing to those steaming from photoemission
and XAS as well as expected from the chemical trends. For instance, a
detailed examination of the giant spin-splitting of free excitons in Ga$_{1-x}$Mn$_x$N
leads to $\beta N_0 = +1.4 \pm 0.3 eV$ \cite{Pacuski:2007_a}, if  the s-d exchange
integral $\alpha N_0 = 0.2 \pm 0.1$~eV is assumed, though
the probed charge state may differ from that examined by photoemission. Similarly, the study of
magnetoluminescence of bound excitons in Zn$_{1-x}$Mn$_x$O implies $|\beta N_0| \approx 0.1$~eV \cite{Przezdziecka:2006_a}.
Interestingly, the contradiction in question is not limited to Mn-based
nitrides and oxides -- it is also rather evident for Zn$_{1-x}$Co$_x$O, where $\beta N_0 = -3.4$~eV according
to XAS and in agreement with the chemical trends \cite{Blinowski:2002_a},
while a thorough investigation of the free exciton splitting results in
$\beta N_0 \approx -0.6$~eV or 1~eV, depending on the assumed ordering
of the valence band subbands \cite{Pacuski:2006_a}.

In this paper we present arguments indicating that nitrides and oxides
belong to an original and unexplored class of DMS, in which the conditions for the strong
coupling between the band hole and the localized magnetic impurity
are met. We then calculate the spectral density $A(\omega)$
of the band-edge states by a generalized alloy theory and show that in the strong
coupling limit $A(\omega)$ exhibits two maxima corresponding
to the bonding and anti-bonding states, respectively. We identify the former as the deep band-gap
hole trap observed optically, while the latter as the position of the band-edge,
and thus of the free exciton, renormalized in a non-trivial way by the hole coupling to the system
of randomly distributed magnetic impurities.
As we demonstrate, our model makes it possible to explain the
reversed sign and the reduced magnitude of the apparent $\beta N_0$, as determined by the free
exciton splitting, as well as the dependence of the energy gap on the magnetic ion
concentration $x$. In this way we reconcile the results of
photoemission, XAS, optical, and magnetooptical studies and show their consistency
with the chemical trends.  Furthermore, our findings provide a new piece
of evidence for the inverse order of the valence subbands in ZnO as well as
reconfirm that the local spin density approximation (LSDA) results in
too high energetic positions of the localized d states.

The  description of exciton splittings in the works referred to above has been carried out
within the time-honored virtual-crystal approximation (VCA) and molecular-field approximation
(MFA), according to which the interaction of carriers and localized spins
leads to bands' splitting proportional to the relevant bare exchange integral and
magnetization of the subsystem of magnetic ions. However, while this approach
describes very well the giant exciton splittings in the tellurides \cite{Gaj:1979_a}
it has already been called into question in the
case of Cd$_{1-x}$Mn$_x$S \cite{Benoit:1992_a}. In this system, unexpected dependencies
of the band-gap and of the apparent exchange integral $\beta^{(app)}$ on $x$ have been
explained by circumventing  VCA and MFA  either employing a non-perturbative Wigner-Seitz-type model \cite{Benoit:1992_a}
or by generalizing to DMS the alloy theory \cite{Tworzydlo:1994_a}. It has been found that the
physics of the problem is governed by the ratio of a characteristic magnitude of the
total magnetic impurity potential $U$
to its critical value $U_{c}$ at which a bound state starts to form. In particular,
the weak coupling regime, where VCA and MFA apply, corresponds to $U/U_{c} \ll 1$.  By modeling
the total potential of the isoelectronic magnetic impurity with a square well of
the radius $b$ one obtains \cite{Benoit:1992_a,Tworzydlo:1994_a},
\begin{equation}
U/U_{c} = 6m^*[W -(S+1)\beta/2]/(\pi^3\hbar^2b),
\end{equation}
where the bare valence band offset $W N_0 = \mbox{d}E_v(x)/\mbox{d} x$,  $S$ is the impurity spin,
and $m^*$ is the effective mass of a particle with the spin $s = 1/2$ assumed to
reside in a simple parabolic band.

In order to evaluate $U/U_{c}$ for specific materials, we adopt $\beta = -0.057$~eVnm$^3$, {\em i.e.},
$\beta N_0 \sim a_0^{-3}$ , as implied by the chemical trends \cite{Dietl:2001_a,Blinowski:2002_a}.
Furthermore, we note that the value of $b$ lies presumably between the anion-cation and cation-cation distance,
$\sqrt 3 a_0/4 \lesssim b \lesssim a_0/\sqrt 2$.
Hence, for Cd$_{1-x}$Mn$_x$Te, where $S =5/2$, $m^*/m_0 = 0.65$ and $W N_0 =-0.63$~eV \cite{Wojtowicz:1996_a} we obtain
$0.20 \lesssim U/U_{c} \lesssim 0.33$, in agreement with the notion
that VCA and MFA can be applied to this system. In the case of Cd$_{1-x}$Mn$_x$S, where $m^*/m_0 = 1.0$ and $W N_0 = 0.5$~eV
\cite{Benoit:1992_a} the coupling strength increases to $0.77 \lesssim U/U_{c} \lesssim 1.25$. Hence,
in agreement with experimental findings, Cd$_{1-x}$Mn$_x$S represents a marginal case, in which
the local spin dependent potential introduced by the magnetic ion is too weak to bind the hole, but
too strong to be described by VCA and MFA \cite{Benoit:1992_a,Tworzydlo:1994_a}.

We argue here that even greater magnitudes of $U/U_{c}$ can be expected for nitrides
and oxides, as $a_0$ is smaller, while both $m^*$ and $W$ tend to be
larger comparing to compounds of heavier anions. We assume $m^* = 1.3m_0$ and, in order to estimate $W$,
we note that if the valence band offset is entirely determined by the p-d hybridization,
$W$ and $\beta$ are related \cite{Benoit:1992_a}.
In the case of Ga$_{1-x}$Mn$_x$N, for which the Hubbard correlation energy for the d electrons
$U^{(eff)} = 10.4$~eV and the position of the d-state with respect to the top of
the valence band, $\epsilon_d^{(eff)} = -5.7$~eV \cite{Hwang:2005_a} (determined so-far with accuracy of about 1~eV),
we arrive to  $0.96 \lesssim U/U_{cr} \lesssim 1.6$. Similarly for Zn$_{1-x}$Mn$_x$O, where
$U^{(eff)} = 9.2$~eV and $\epsilon_d^{(eff)} = -1.5$~eV \cite{Okabayashi:2004_a},
we find  $2.0 \lesssim U/U_{c} \lesssim 3.3$. Interestingly, the above evaluation
of $U/U_{c}$ for particular materials remains valid for DMS and DMO
containing other transition metals than Mn in the $2+$ state as, to a first approximation,
$\beta S$ does not depend on $S$ \cite{Blinowski:2002_a}.
Although the quoted values of $U/U_{c}$ are subject of uncertainty stemming from the limited
accuracy of the input parameters as well as from the approximate treatment of the hole
band structure, we are in the position to conclude that magnetically doped nitrides and oxides
are in the strong coupling regime,  $U/U_{c}>1$, where the hole states cannot be described by VCA and MFA.

We evaluate the effect of magnetic impurities upon a single band-edge particle ($k=0$) with the spin $s =1/2$
by employing a generalized alloy theory developed by Tworzyd{\l}o \cite{Tworzydlo:1994_a}. The theory
is built for non-interacting, randomly distributed, and fluctuating  quantum spins ${\bm S}$
characterized by the field-induced averaged polarization  $\langle S_z(T,H)\rangle$.
The potential of the individual impurities is modeled \cite{Benoit:1992_a} by the
square-well potential containing both spin-dependent (exchange) and spin-independent
(chemical shift) contributions, as discussed above. The particle self-energy $\Sigma_{s_z}(\omega)$ is derived
from the Matsubara formalism by summing up an infinite series for the irreducible self-energy in the average $t$-matrix
approximation (ATA) \cite{Tworzydlo:1994_a}. Because of its general applicability for various alloys, we
write down the form of $\Sigma_{s_z}(\omega)$ explicitly, neglecting the direct effect of the magnetic field
on the band states,
\begin{widetext}
\begin{equation}
\Sigma_{s_z}(\omega) = [(S+1+2s_z\langle S_z\rangle)\Sigma_0(S) +(S-2s_z\langle S_z\rangle)\Sigma_0(-S-1)]/(2S+1),
\end{equation}
where $\Sigma_0(J)=  x N_0(v U - 16\mbox{i}\pi m^*UU^{\prime} I\kappa^{\prime}b^5/\hbar^2);
v = 4\pi b^3/3$; $U = -(W + J\beta/2)/v$;  $U^{\prime} = U - E_{vc}/(vN_0)$;\vspace{2mm} \\
\noindent
$E_{vc} = -xN_0(s_z\langle S_z\rangle\beta + W)$;
$\Delta = \mbox{i}(\kappa-\kappa^{\prime})\exp(\mbox{i}\kappa^{\prime})/(\kappa\sin\kappa^{\prime}+\mbox{i}\kappa^{\prime}\cos\kappa^{\prime})$;
$\kappa = b(2m^*\tilde{E}/\hbar^2)^{1/2}$; $\kappa^{\prime} = b[2m^*(\tilde{E}-U)/\hbar^2]^{1/2}$;
\begin{equation}
I = \{3(1+\Delta) + 3\kappa^{\prime 2}(1+ \Delta) + 2\mbox{i}\kappa^{\prime 3} +
       3[(\mbox{i}+\kappa^{\prime})^2 + \Delta(\kappa^{\prime 2}-1)]\cos(2\kappa^{\prime})
      +3\mbox{i}[\kappa^{\prime}(2\mbox{i}(1+ \Delta) + \kappa^{\prime})-1]\sin(2\kappa^{\prime})\}/(12\kappa^{\prime 6}).
 \end{equation}
 \end{widetext}
Here  $\tilde{E} = \omega + E_{vc} + \mbox{i}\gamma$,
where $\gamma$ is an intrinsic life-time broadening of the $k=0$ state,
taken here to be 3~meV.

The quantity of interest is the spectral density of states,
\begin{equation}
A_{s_z}(\omega) = -\frac{1}{\pi}\mbox{Im}\frac{1}{\omega +  \mbox{i}\gamma - \Sigma_{s_z}(\omega)},
\end{equation}
whose maxima provide the position of the band edge $\tilde{E_0}$ in the presence of spin and chemical disorder as a function
of $\langle S_z(T,H)\rangle$. Most of relevant experiments for magnetically doped nitrides and oxides has been
carried out at low concentrations $x \lesssim 3$\%, where effects of interactions among localized spins, neglected
in the present approach, are not yet important. In this case, $\langle S_z(T,H)\rangle$ can
be determined from magnetization measurements or from the the partition function of spin hamiltonian
for the relevant magnetic ion at given temperature $T$ and the magnetic field $H$.  For higher $x$, it is tempting to
take into account effects of the short-range antiferromagnetic superexchange according
to the standard recipe  \cite{Gaj:1979_a,Dietl:2001_a}, {\em i.~e.}, via replacement of $x$ by $x_{eff}$ and $T$ by $T+T_{AF}$,
where $T_{AF}(x,T) > 0$ and $x_{eff}(x,T) < x$. However, in the strong coupling case this procedure is rather inaccurate as
aniferromagnetically aligned pairs contribute also to the band-edge shift.

\begin{figure}
\includegraphics[scale=0.35]{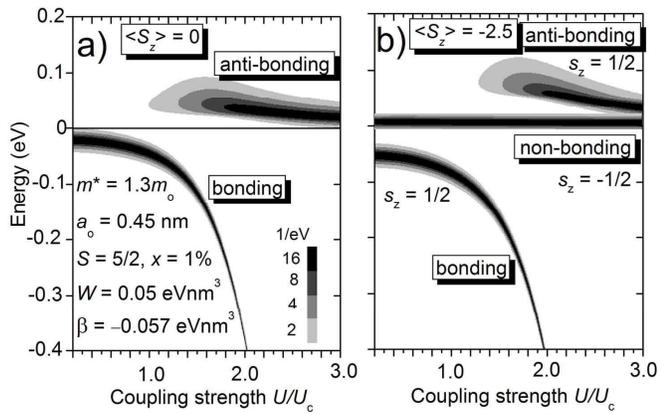}
\caption{Energy distribution (grey scale) of the spectral density $A_{s_z}(\omega)$ at the band-edge  ($k=0$) as a function of the strength of
the coupling between the spin $1/2$ particle and the system of unpolarized (a) and spin-polarized (b) magnetic impurities. The magnitude
of $U/U_{c}$ is changed by the range of the impurities' potential $b$.}
 \label{fig:spectrum_mp}
\end{figure}

Figure 1(a) presents a grey scale plot of the spectral density $A_{s_z}(\omega)$ as a function of the coupling strength in
the absence of spin polarization,  $\langle S_z\rangle =0$, so that the spin degeneracy is conserved, $A_{1/2}(\omega)
= A_{-1/2}(\omega)$. In the weak coupling limit, $U/U_c \ll 1$, the band-edge position $\tilde{E_0}$ is obviously given by $-xN_0W$.
For higher $U/U_c$ a downward shift of $\tilde{E_0}$ is visible, particularly rapid for $U/U_c >1$, when the magnetic ion potential
is strong enough to bind a hole, even if the magnetic ion is an isoelectronic impurity. Such an effect is also expected within the dynamic mean-field
approximation \cite{Chattopadhyay:2001_a}.
The small magnetic polaron formed in this way, reminiscent of the Zhang-Rice singlet, will be built also
of $k$ states away from the Brillouin zone center, as discussed previously \cite{Dietl:2002_a}.  This band-gap hole
trap should be visible in photoionization experiments d$^5\rightarrow$ d$^5$ + h + e, as indeed observed in both  n-Ga$_{1-x}$Mn$_x$N
\cite{Wolos:2004_a}
and Zn$_{1-x}$Mn$_x$O \cite{Kittilstved:2006_a} but assigned in those works to  d$^5\rightarrow$ d$^4$ + e transitions. In Ga$_{1-x}$Mn$_x$N samples, in which
donors were partly compensated, intra-center excitations corresponding to this level were also detected at 1.4~eV \cite{Korotkov:2002_a}
and analyzed in considerable details \cite{Wolos:2004_a,Wolos:2004_b,Marcet:2006_a}. While symmetry considerations cannot discriminate between
d$^5$ + h and d$^4$ many-electron configurations, the larger crystal-field splitting and the smaller Huang-Rhys factor
in Ga$_{1-x}$Mn$_x$N comparing to their values in (II,Cr)VI compounds \cite{Wolos:2004_b,Marcet:2006_a} point to a relatively large localization radius,
expected for the d$^5$ + h case. Our interpretation reconfirms  also the limited accuracy of {\em ab initio} computations within LSDA
in the case of transition metal insulators, which place the d$^4$ Mn level within the band-gap of GaN
\cite{Titov:2005_a}.

\begin{figure}
\includegraphics[scale=0.36]{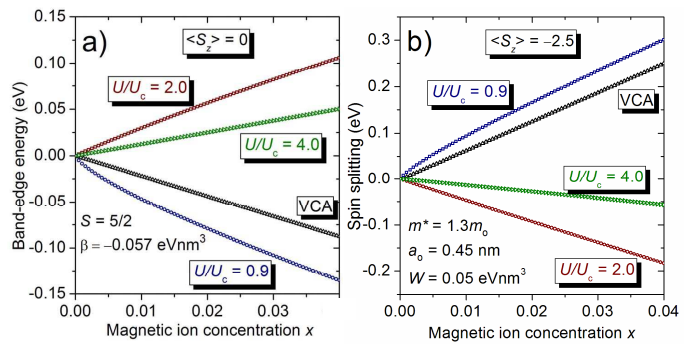}
\caption{Dependence of the band-edge position $\tilde{E_0}$(a) and of its spin splitting $\Delta_s$ (b) on the concentration of magnetic impurities $x$ for
three values of the coupling strength in comparison to the values expected within VCA and MFA.}
 \label{fig:vs_x}
\end{figure}

Interestingly, as seen in Fig.~1(a), when $U/U_c$ increases beyond 1, the spectral density is gradually transferred from the bonding
state discussed above to an anti-bonding
state appearing above the band-edge energy $E_0$ expected within VCA and MFA.  We identify this state with
the actual valence band-edge for $U/U_c>1$, whose position $\tilde{E_0}$ determines, {\em e.~g.},
the onset of interband optical transitions and the free exciton energy.
In Fig.~2(a), we plot $\tilde{E_0}(x)$ for $U/U_c < 1$ and $U/U_c > 1$ in comparison to the VCA and MFA expectations.
We see that if $U/U_c < 1$, the correction to VCA and MFA leads to a reduction of the band-gap with $x$, as observed
in Cd$_{1-x}$Mn$_x$S \cite{Tworzydlo:1994_a},  Zn$_{1-x}$Mn$_x$Se \cite{Bylsma:1986_a}, and  Ga$_{1-x}$Mn$_x$As \cite{Dietl:2001_a}.
In contrast, for $U/U_c > 1$ the present model predicts an increase of the gap with $x$, in agreement with the data
for  Ga$_{1-x}$Mn$_x$N \cite{Marcet:2006_a}, Zn$_{1-x}$Mn$_x$O \cite{Fukumura:1999_a}, and Zn$_{1-x}$Co$_x$O \cite{Pacuski:2006_a}.

\begin{figure}
\includegraphics[scale=0.36]{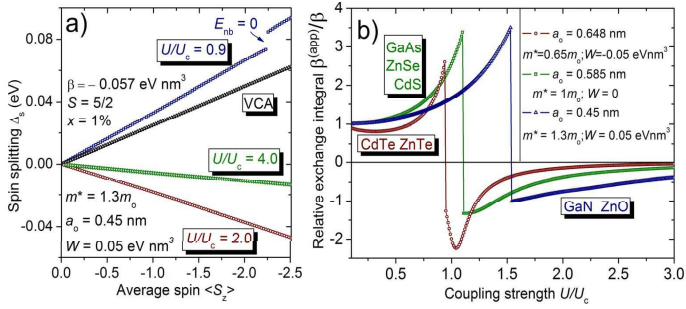}
\caption{Dependence of the band-edge spin splitting on the spin polarization,  $\Delta_s(\langle S_z\rangle)$  for
three values of the coupling strength in comparison to the values expected within VCA and MFA, $\Delta_s^{VCA}(\langle S_z\rangle)$(a).
Normalized apparent p-d exchange integral defined as $\Delta_s(-2.5)/\Delta_s^{VCA}(-2.5)$ for various combinations of the materials parameters (b) }
 \label{fig:delta_beta}
\end{figure}

As shown in Fig.~1(b), the saturating magnetic field, such that $\langle S_z(T,H)\rangle = -2.5$, produces a downward and upward shift of the
bonding and anti-bonding states (which correspond to $s_z = 1/2$), respectively, and leads to the appearance of a non-bonding state ($s_z = -1/2$),
whose energy is virtually independent of $U/U_c $. Remarkably, the latter resides below the anti-bonding level, which means that
the character of the apparent p-d exchange changes from antiferromagnetic for $U/U_c < 1$ to ferromagnetic for $U/U_c > 1$.
Extensive studies of the giant free excitons' splitting $\Delta_s$
in Ga$_{1-x}$Mn$_x$N \cite{Pacuski:2007_a} and Zn$_{1-x}$Co$_x$O \cite{Pacuski:2006_a} demonstrated a linear
dependence of $\Delta_s$ on  $x$ and $\langle S_z(T,H)\rangle$, as expected  within VCA and MFA. Figures 2(b) and 3(a) demonstrate that approximately
linear dependence of $\Delta_s$ on $x$ and $\langle S_z(T,H)\rangle$ is predicted within the present theory, too, except
for an anticrossing behavior occurring when the non-bonding state is in a resonance with the non-normalized band-edge, $E_0 =0$.
Accordingly, for the
sake of comparison to the experimental determinations,
$\Delta_s$ can be characterized by an apparent p-d exchange integral according to $\Delta_s = -xN_0\beta^{(app)}\langle S_z(T,H)\rangle$.
In ~Fig.~3(b), we depict the expected evolution of  $\beta^{(app)}/\beta$ with $U/U_c$ for several values of $W$ and $m^*$. We see
that our theory explains the sign reversal of  $\beta^{(app)}$ and its reduced amplitude comparing to the $\beta$ values determined
by the photoemission and XAS experiments. Our findings may imply also that the exchange coupling between the band hole and the hole on the
d$^5$ + h center is not qualitatively important in Ga$_{1-x}$Mn$_x$N as well as verify
the reversed order of valence band subbands in ZnO \cite{Lambrecht:2002_a}.

In summary, the approach put forward here allows one to reconcile the findings of photoemission, XAS, optical, and magnetooptical
measurements accumulated over recent years for the magnetically doped nitrides and oxides.
In view of our results, these systems form an outstanding class of materials, in which a number of concepts developed earlier for DMS
has to be revised. In particular,  the strong hole localization discussed here, along with the issues of self-compensation
and solubility, need to be overcame in order to observe the hole-mediated ferromagnetism in nitrides and oxides.

This work was supported in part by and by the EC project NANOSPIN (FP6-2002-IST-015728).
I would like to thank A. Bonanni, F. Matsukura, H. Ohno, and W. Pacuski for valuable
discussions.

\end{document}